# Spin-Density-Driven Hybrid Spin Splitting in Collinear Compensated Magnets

Peibo Xu,[1,3] Yixuan Che,[4] Yuxuan Li,[1] Haifeng Lv,[2,*] Xiaojun Wu,[2,3,*] and Jinlong Yang[1,2,3]

[1]*School of Emerging Technology and Hefei National Research Center for Physical Sciences at the Microscale, University of Science and Technology of China, Hefei, Anhui 230026, China*

[2]*State Key Laboratory of Precision and Intelligent Chemistry, School of Chemistry and Materials Science, and Collaborative Innovation Center of Chemistry for Energy Materials (iChEM), University of Science and Technology of China, Hefei, Anhui 230026, China*

[3]*Hefei National Laboratory, University of Science and Technology of China, Hefei, Anhui 230088, China*

[4]*Chair of Theoretical Chemistry, Technische Universität Dresden, 01062 Dresden, Germany*

[*]Contact authors: xjwu@ustc.edu.cn (X.W.); hflv@ustc.edu.cn (H.L.)

Nonrelativistic spin splitting in collinear compensated magnets is constrained by spin-group symmetry, with distinct alternating and Zeeman-type splitting typically associated with altermagnets and fully compensated ferrimagnets, respectively. From a spin-density perspective, we propose a class of hybrid compensated magnets in which alternating-like and Zeeman-type spin splitting coexist within a single collinear compensated state. Using a minimal two-sector model, we demonstrate that this coexistence arises from competition between directional sublattice dispersion and the sector-dependent energy biases arising from the same structural inequivalence. First-principles calculations provide a proof of concept in a crossed bilayer of unequal-width Janus graphene nanoribbons, where the structurally inequivalent ribbons retain approximately symmetry-related spin-density distributions. The valence and conduction sectors support direction-selective spin transport and energy-selective spin-polarized states, respectively. These results connect the spatial organization of spin density with the electronic response to structural inequivalence, providing a design principle for hybrid spin splitting in compensated magnets.

*Introduction*—Unconventional collinear compensated magnets combine the merits of ferromagnets and antiferromagnets, featuring both vanishing net magnetization and nonrelativistic spin splitting (NRSS). Represented by altermagnets (AMs) and fully compensated ferrimagnets (CFIMs) [1–8], unconventional compensated magnetism supports a rich variety of phenomena including magneto-optical effects [9–11], Hall responses [12–14], spin-dependent transport [15–21], and superconductivity [22–24].

While both belong to compensated magnets regime, AMs and CFIMs exhibit fundamentally distinct spin-band structures. In altermagnets, a rotation or mirror operation connecting the opposite-spin sublattices enforce a momentum-dependent spin splitting that changes sign between symmetry-related directions [1,2,25–28]. However, the absence of such a sublattice-connecting operation in fully compensated ferrimagnets permits a typical Zeeman-type splitting across the Brillouin zone [3,4,6]. Because these disparate splitting patterns stem from incompatible global symmetry conditions, AMs and CFIMs have been investigated as separate magnetic classes [4,6]. Unifying these two contrasting spin-splitting patterns within a single compensated system would unlock multifaceted, tunable spin transport free from stray fields. However, the principles for realizing the coexistence of these contrasting splitting patterns in a single compensated magnetic state remain to be clarified.

Here, we propose a class of hybrid compensated magnets (HCMs), hosting coexisting alternating-like and Zeeman-type spin splitting at zero net magnetization. We construct a minimal two-sector model to demonstrate that this coexistence emerges when distinct electronic subspaces experience different effective energy biases from the same structural inequivalence. First-principles calculations on two-dimensional crossed nanoribbons provide a proof of concept, with the valence and conduction sectors supporting direction-selective spin transport and energy-selective spin-polarized states, respectively.

*Structural and spin density perspective*—In the absence of SOC, nonrelativistic spin group theory combines independent spin and real-space transformations into operations $[g_s||g]$, providing a systematic symmetry framework for classifying magnetic orders [29,30]. For altermagnets, the opposite-spin sublattices are related by a $[C_2||A]$ symmetry operation, where $A$ is a rotation or reflection in real lattice space. This structural correspondence between the sublattices guarantees a matching $[C_2||A]$ relation in the real-space spin density [Fig. 1(a)], enforcing alternating spin splitting in momentum space. Conversely, in typical compensated ferrimagnets, the opposite-spin sublattices are structurally inequivalent and not related by any crystal symmetry. Their spin densities generally reflect this inequivalence, allowing global Zeeman-type spin splitting [Fig. 1(b)].

While symmetry analysis identifies the allowed spin-splitting patterns, it does not by itself quantify how strongly different electronic subspaces respond to a given structural inequivalence. Crucially, even when opposite-spin sublattices are structurally inequivalent, their spin-density distributions may preserve an approximate correspondence under $[C_2||A]$ [Fig. 1(c)]. Electronic subspaces with different spatial characters can respond differently to this structural asymmetry. We therefore propose a HCM regime in which alternating-like and Zeeman-type spin splitting coexist within a single compensated magnet.

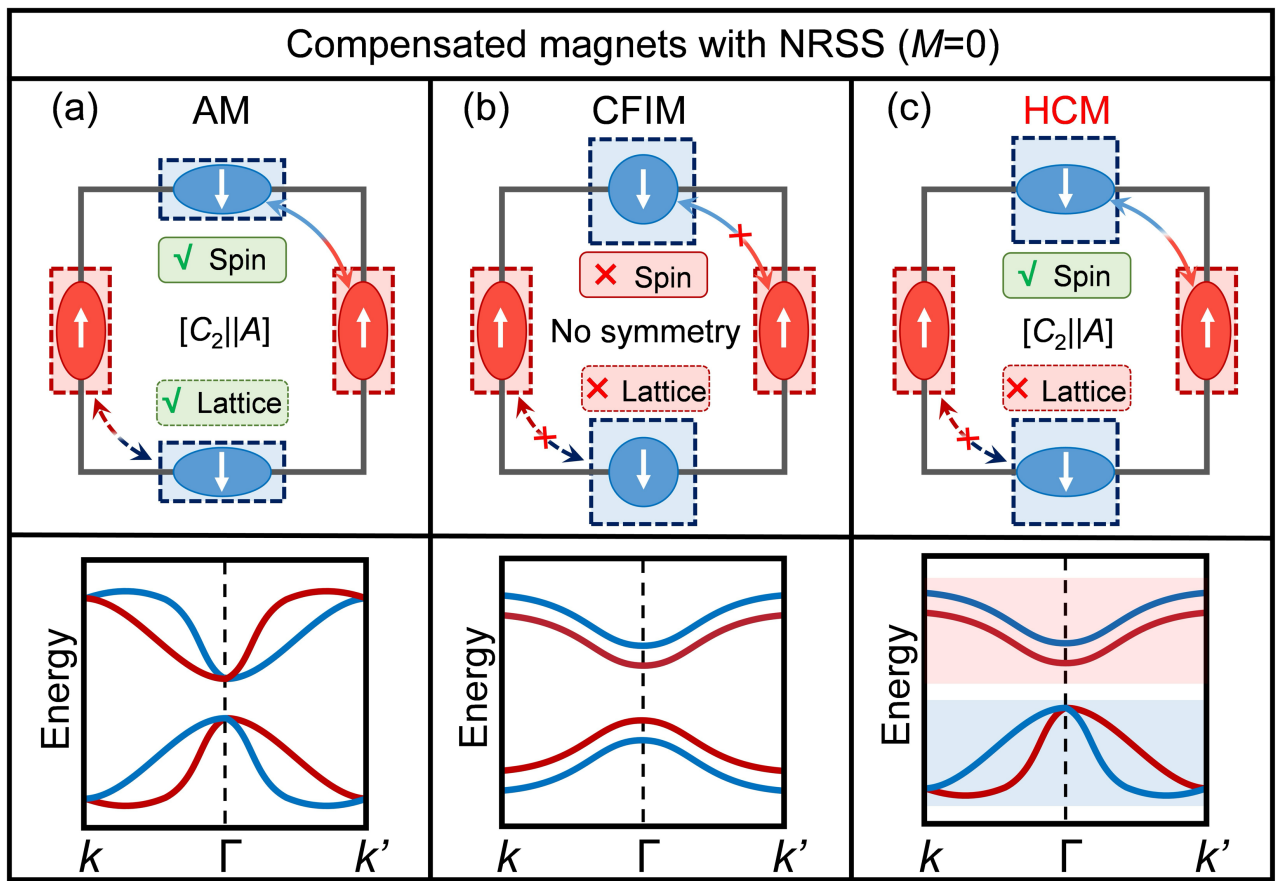

FIG. 1. Symmetry and band-splitting paradigms in compensated magnets with NRSS. The red and blue ellipses represent opposite-spin density distributions, while the red and blue dashed boxes identify the corresponding opposite-spin sublattices. Solid arrows at the upper right and dashed arrows at the lower left schematically indicate symmetry relations between the density distributions and between the sublattices, respectively. (a) AM: the spin-group operation $[C_2\|A]$ maps both the opposite-spin sublattices and their spin-resolved densities, enforcing an alternating splitting. (b) CFIM: no spatial operation connects the structurally inequivalent opposite-spin sublattices, thereby allowing a global Zeeman-type splitting. (c) HCM: the sublattices remain structurally inequivalent, so $[C_2\|A]$ is not a symmetry of the full system, whereas their opposite-spin density distributions retain an approximate $[C_2\|A]$ related correspondence. The resulting subspace-dependent responses to structural inequivalence produce alternating-like and Zeeman-type spin-splitting patterns in the respective shaded energy regions.

To elucidate the impact of structural inequivalence on the electronic structure, we construct a minimal two-sector model ($\alpha$ = 1, 2) for the coexistence of alternating and Zeeman-type spin splittings. For conserved spin $s = \pm 1$, the spin-resolved Hamiltonian reads:

$$\hat{H} = \sum_{\mathbf{k},s} \hat{\Psi}_{\mathbf{k}s}^{\dagger} H_s(\mathbf{k}) \hat{\Psi}_{\mathbf{k}s}, \qquad H_s(\mathbf{k}) = \begin{pmatrix} h_{1s}(\mathbf{k}) & V_{12}\tau_0 \\ V_{12}\tau_0 & h_{2s}(\mathbf{k}) \end{pmatrix},$$

$$h_{\alpha s}(\mathbf{k}) = \varepsilon_\alpha \tau_0 + t_\alpha(\mathbf{k})\tau_x + [a_\alpha f(\mathbf{k}) + \delta_\alpha + sJ_\alpha]\tau_z. \quad (1)$$

Here, $\hat{\Psi}_{\mathbf{k}s} = (c_{A1}, c_{B1}, c_{A2}, c_{B2})_{\mathbf{k}s}^T$, where $A$ and $B$ denote opposite-spin sublattices. The Pauli matrices $\tau_{x,z}$ act in the sublattice space, and $\tau_0$ is the identity matrix. In each sector, $\varepsilon_\alpha$ sets the band center, $t_\alpha(\mathbf{k})$ describes spin-independent intersublattice hopping, and $sJ_\alpha$ represents the staggered exchange field. The form factor satisfies $f(A\mathbf{k}) = -f(\mathbf{k})$ under a real-space rotation or reflection $A$, such that $a_\alpha f(\mathbf{k})$ describes a direction-dependent sublattice dispersion whose interplay with $sJ_\alpha$ produces the alternating component of the spin splitting. The two sectors are weakly hybridized by $V_{12}$. Crucially, structural inequivalence enters the model through sector-dependent sublattice biases $\delta_\alpha$, which arise from different projections of the same structural perturbation onto the two electronic sectors.

For the weakly hybridized sectors considered here, the contrasting splitting textures are governed primarily by the competition between directional sublattice dispersion and this structural bias. When $|\delta_1| \ll \max_{\mathbf{k}} |a_1 f(\mathbf{k})|$, sector 1 (S1) retains a predominantly alternating-like splitting $\Delta_1(A\mathbf{k}) \simeq -\Delta_1(\mathbf{k})$. By contrast, when $|\delta_2| > \max_{\mathbf{k}} |a_2 f(\mathbf{k})|$, structural inequivalence fixes the sign of the splitting throughout sector 2 (S2), producing a same-sign Zeeman-type spin splitting. For a concrete illustration of the two spin-splitting textures, we consider a representative square-lattice model. Taking the lattice constant as unity, we choose the $d_{x^2-y^2}$-type form factor

$$f(\mathbf{k}) = \cos k_x - \cos k_y, \quad (2)$$

and the intersublattice hopping

$$t_\alpha(\mathbf{k}) = 4t_\alpha \cos\frac{k_x}{2}\cos\frac{k_y}{2}. \quad (3)$$

To isolate the role of structural inequivalence, we set $t_1=t_2$, $a_1=a_2$, and $J_1=J_2$. The sector-dependent energies $\varepsilon_\alpha$ separate S1 and S2 spectrally, whereas $\delta_\alpha$ controls their contrasting splitting patterns. Figure 2(a) shows both sectors in the same spectrum. The conduction-band sector S2 exhibits same-sign spin splitting within each band branch throughout the Brillouin zone, with the lower-branch splitting $\Delta E_n(k) = E_{n\uparrow}(k) - E_{n\downarrow}(k)$ ranging from -2.00$t$ to -1.04$t$ [Fig. 2(b)]. By contrast, the valence-band sector S1 exhibits predominantly alternating-like spin splitting. Its upper-branch splitting ranges from -1.30$t$ to 1.50$t$ and reverses sign between rotation-related momentum regions, forming a $d$-wave-like texture [Fig. 2(c)]. For the parameters used in Fig. 2, both spin channels share a finite band gap with two fully occupied bands in each channel, yielding $N\uparrow = N\downarrow$ and hence a compensated magnetic state. Further details of the model analysis are provided in Note S1.

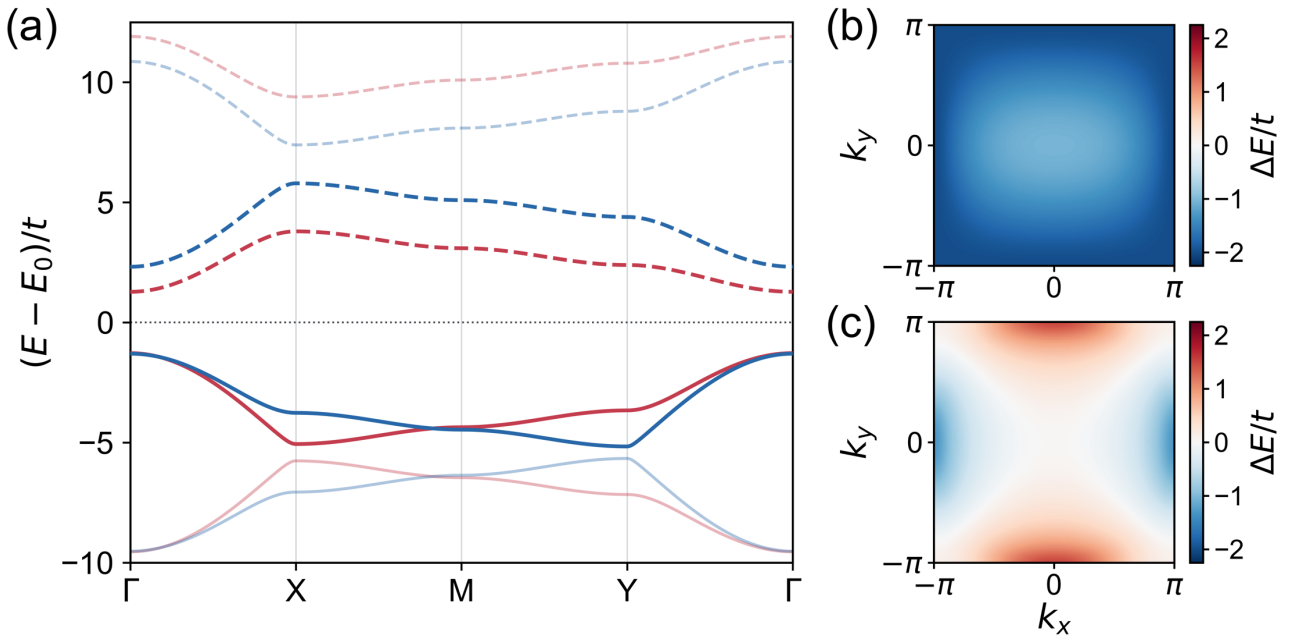

FIG. 2. Coexisting spin-splitting patterns in the minimal model. (a) Spin-resolved band structure along the square-model path Γ-X-M-Y-Γ (Fig. S1). Energies are measured from the model midgap $E_0$ in units of $t$. Red and blue denote spin-up and spin-down states, while solid and dashed lines represent sectors 1 and 2, respectively. (b) Spin splitting of the S2 lower branch, corresponding to the highlighted dashed band pair in (a), with the same sign throughout the model Brillouin zone. (c) Spin splitting of the S1 upper branch, corresponding to the highlighted solid band pair in (a), with a sign reversal between rotation-related momentum regions. Both maps use the same color scale. The parameters of sectors 1 and 2 are $\{\varepsilon_1, t_1, a_1, J_1, \delta_1\} =$

$\{-6.00, 1.00, 0.35, 1.00, 0.05\}$ , $\{\varepsilon_2, t_2, a_2, J_2, \delta_2\} = \{6.00, 1.00, 0.35, 1.00, 2.5\}$ and $V_{12} = 0.15$.

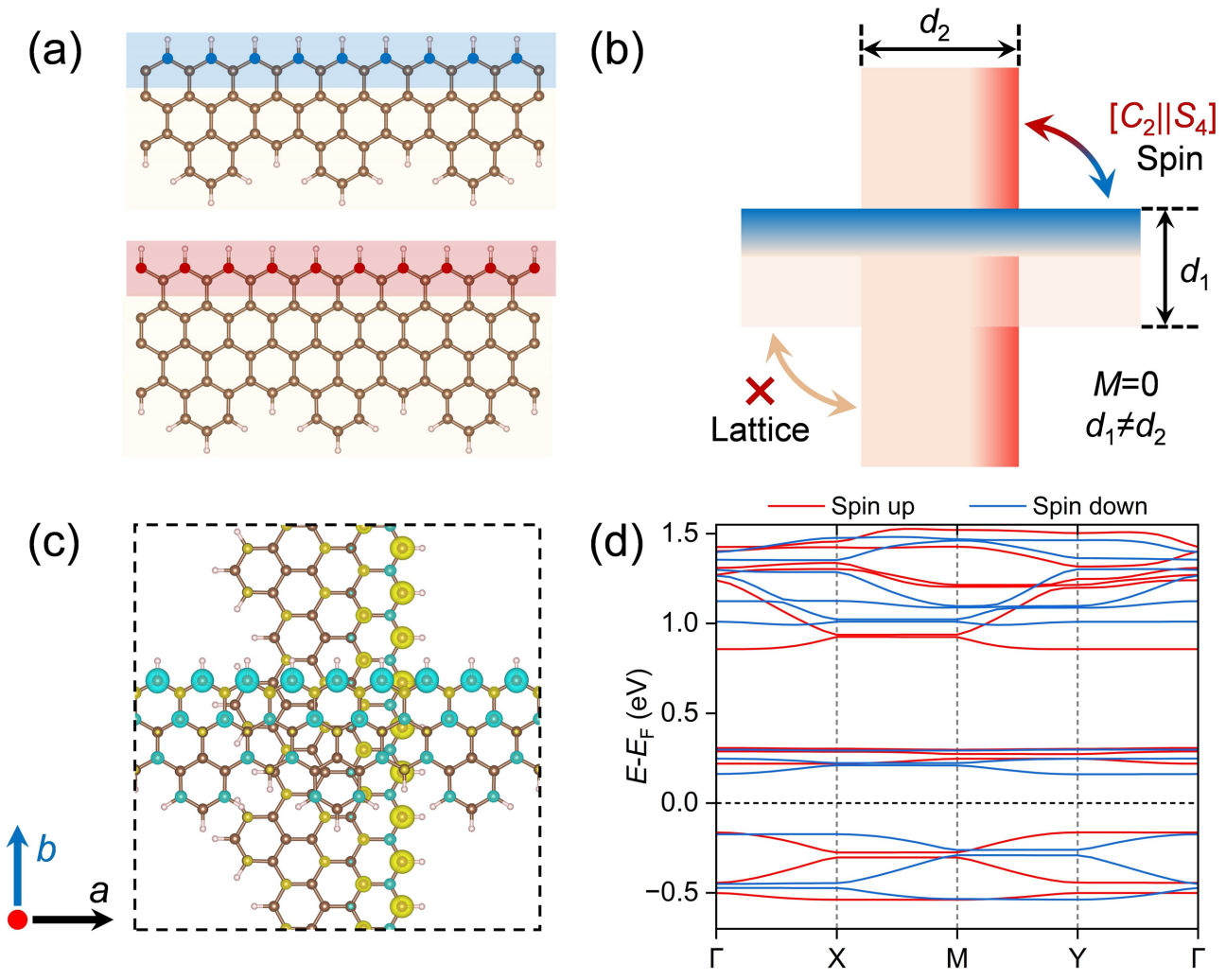

FIG. 3. Crossed-nanoribbon realization of coexisting spin-splitting patterns. (a) Unequal-width graphene nanoribbons with single-edge-dominated magnetism. Blue and red denote opposite signs of the spin density, which is predominantly localized along the zigzag edges. (b) Schematic of the crossed-ribbon realization. The width mismatch $d_1 \neq d_2$ breaks the lattice-level relation between the opposite-spin sublattices, while their spin-density distributions remain approximately related by $[C_2||S_4]$. The antiferromagnetic interlayer coupling produces a compensated state with $M$=0. (c) Top view of the crossed nanoribbon unit cell and the corresponding opposite-sign spin-density isosurfaces at an isovalue of 0.005 e bohr$^{-3}$. (d) Nonrelativistic spin-resolved band structure along Γ-X-M-Y-Γ. Red and blue represent the spin-up and spin-down channels, respectively, and the Fermi level is set to zero.

*First-principles realization in crossed nanoribbons*— The model therefore suggests a design principle for hybrid spin splitting: introduce structural inequivalence between opposite-spin sublattices while retaining approximately symmetry-related spin-density distributions. Hybrid spin splitting can arise when different electronic sectors respond differently to this structural inequivalence. Janus graphene nanoribbons with magnetism confined to a single zigzag edge provide a natural platform for this design [31,32]. These ribbons have recently been synthesized experimentally, with the magnetic states at the modified edge quenched and the spin density predominantly localized along the opposite pristine zigzag edge [33]. As shown in Fig. 3(a), this edge localization allows ribbons of different widths to maintain nearly identical local spin-density distributions. Furthermore, graphene nanoribbons can be stacked into van der Waals bilayers [34], providing the layer degree of freedom for engineering the symmetry relations between opposite-spin sublattices.

Therefore, we construct a crossed bilayer of (3,2) and (4,2) Janus graphene nanoribbons with nearly equal in-plane lattice constants [Figs. 3(a)-(c) and Table S1]. In these ribbon labels, the first index specifies the parent ribbon width in terms of zigzag carbon chains, while the second indicates the spacing of additional benzene rings along the modified edge [33]. Consequently, the width mismatch $d_1 \neq d_2$ removes the $S_4$ relation between the complete lattices, while their spin-density distributions remain approximately connected by $[C_2||S_4]$ operation, which has been adopted to construct two-dimensional altermagnets in bilayer systems [35,36]. This combination of global structural inequivalence and locally related spin-density distributions provides a natural platform to validate the proposed coexistence mechanism. Crucially, the two layers are coupled antiferromagnetically, with total magnetic moments exactly compensated [Fig. 3(c) and Table S1]. The corresponding nonrelativistic band structure exhibits both predicted splitting patterns. The highest valence band exhibits alternating-like spin splitting, with $\Delta E$ changing from +97 meV at Y point to -102 meV at X point. By contrast, the higher conduction band near 1.0 eV maintains the same splitting sign throughout the plotted path, with $\Delta E$ ranging from -153 to -85 meV, consistent with a Zeeman-type splitting.

The spin-resolved partial charge densities reveal contrasting distributions over the layer-separated sublattices for the two splitting sectors. In the alternating-like sector, selected opposite-spin states of the highest valence band at X and Y are predominantly localized on different ribbons and exhibit an approximate rotational correspondence [Fig. S2], suggesting a relatively small effective sublattice bias. In the Zeeman-type sector, the opposite-spin states of the selected conduction band are predominantly localized on the same layer [Fig. S3], indicating a common layer preference consistent with a larger effective sublattice bias. Together with the momentum dependence of the spin splitting, these spatial features support the physical picture provided by the effective model presented above.

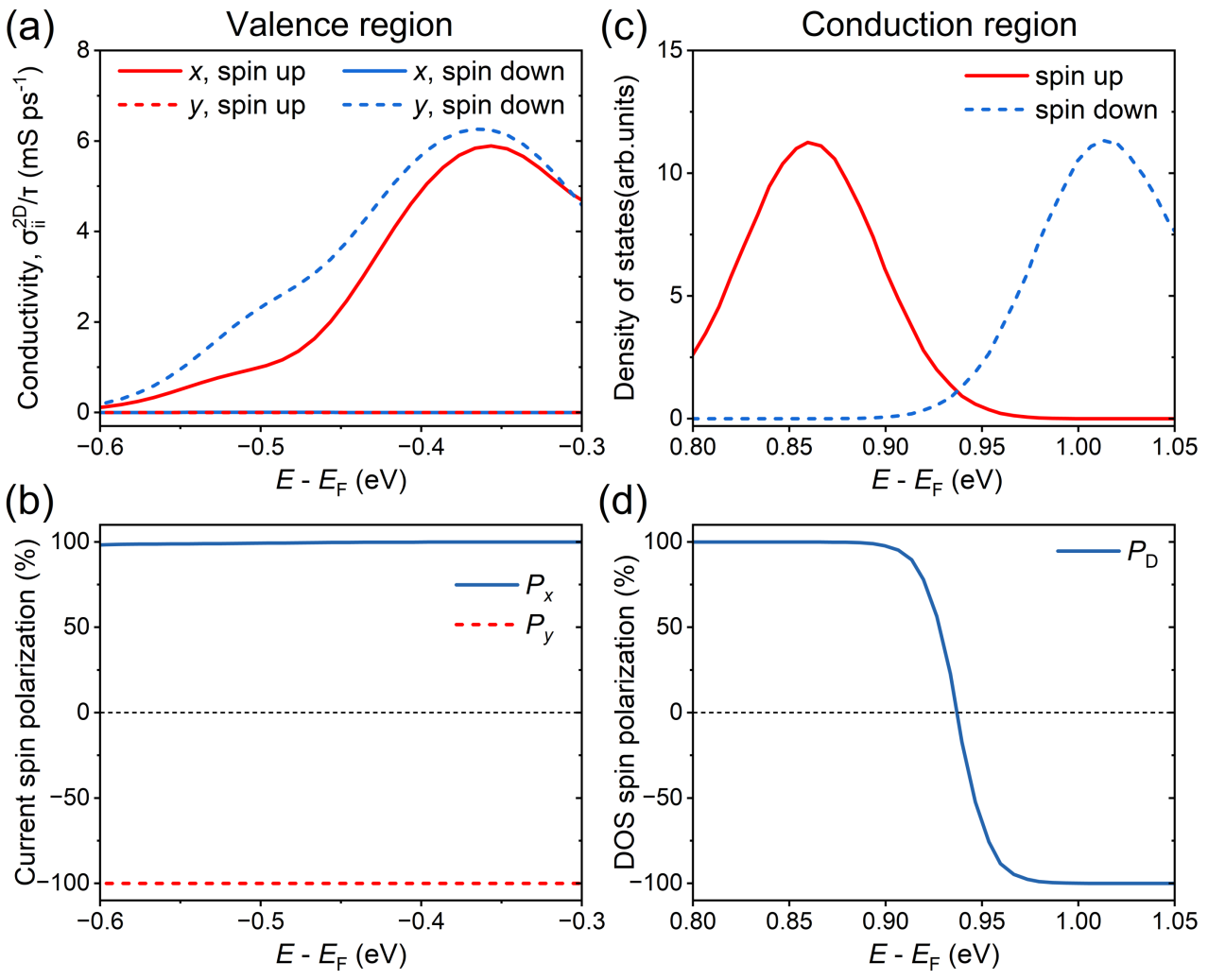

FIG. 4. Complementary spin responses associated with coexisting spin-splitting patterns. (a) Spin-resolved sheet conductivities per relaxation time in the valence region with sign-changing spin splitting. Solid and dashed curves denote transport along $x$ and $y$, respectively. (b) Corresponding current spin polarizations $P_x$ and $P_y$. (c) Spin-resolved DOS in the conduction region with same-sign spin splitting. Red and blue curves indicate spin-up and spin-down channels, respectively. (d) Corresponding DOS spin polarization $P_D$.

The coexisting spin-splitting patterns offer complementary routes to spin selectivity within a single compensated magnet. To illustrate their functional responses, spin-resolved longitudinal conductivities are calculated at 300 K within the rigid-band approximation (see Note S2 and Fig. S4). The longitudinal current spin polarization is defined as

$$P_i(E) = \frac{\sigma_{ii}^{\uparrow}(E) - \sigma_{ii}^{\downarrow}(E)}{\sigma_{ii}^{\uparrow}(E) + \sigma_{ii}^{\downarrow}(E)}, \qquad i = x, y. \tag{4}$$

Here, $\sigma_{ii}^{\uparrow,\downarrow}$ are the spin-resolved longitudinal conductivities for an electric field applied along direction $i$. In the valence region, the sign-changing splitting is accompanied by strong directional spin selectivity: spin-up carriers dominate conduction along $x$, whereas spin-down carriers dominate along $y$ [Fig. 4(a)]. The two dominant sheet conductivities per relaxation time exhibit similar energy dependences, with comparable peaks of approximately 6 mS $ps^{-1}$ near -0.36 eV, while the spin-down conductivity along $x$ and the spin-up conductivity along $y$ remain close to zero. This correspondence further corroborates the altermagnetic-like character of the valence-band sector. The resulting current spin polarizations $P_x \approx +100\%$ and $P_y \approx -100\%$ in the valence-band sector [Fig. 4(b)] reveal a pronounced altermagnetic-like transport response, with opposite spins dominating conduction along the two orthogonal directions.

In contrast, the conduction region with same-sign spin splitting exhibits separated peaks in the spin-resolved density of states (DOS) [Fig. 4(c)]. The dominant spin character switches from spin up to spin down as energy increases. The DOS spin polarization is defined as,

$$P_D = \frac{D_{\uparrow} - D_{\downarrow}}{D_{\uparrow} + D_{\downarrow}}, \tag{5}$$

where $D\uparrow$ and $D\downarrow$ denote the spin-resolved DOS. As energy increases from 0.80 to 1.05 eV, $P_D$ changes from nearly +100% to nearly -100%, reflecting a switch from spin-up to spin-down dominance. Together, these results highlight the complementary functions of the coexisting splitting patterns: direction-selective spin transport in the valence sector and energy-selective spin-polarized states in the conduction sector within a single compensated magnet.

*Conclusion*—In summary, we propose a class of hybrid compensated magnet hosting coexisting nonrelativistic spin-splitting patterns characteristic of altermagnets and fully compensated ferrimagnets in different energy sectors. A minimal two-sector model describes the same structural inequivalence through sector-dependent sublattice biases. Competition between these biases and directional sublattice dispersion allows alternating-like and Zeeman-type spin splitting to coexist. First-principles calculations on a crossed bilayer of unequal-width Janus graphene nanoribbons provide a concrete realization, with alternating-like valence-band splitting and same-sign conduction-band splitting at zero net magnetization. These sectors support direction-selective spin transport and energy-selective spin-polarized states, respectively. Our results provide a design strategy for combining distinct spin-splitting patterns and complementary spin functions within a single compensated magnetic state.

*Acknowledgements*—The research is supported by the National Natural Science Foundation of China (Grants Nos. 22225301, 22303092, and 22321001), the Quantum Science and Technology-National Science and Technology Major Project (Grant No. 2021ZD0303302), the Fundamental Research Funds for the Central Universities (Grants Nos. 20720250005, WK2490000001, and WK2490000002), the National Advanced Talent Cultivation Center for Chemistry of USTC, the open research fund of Key Laboratory of Precision and Intelligent Chemistry, the robotic AI-Scientist platform of Chinese Academy of Sciences, and the Super Computer Center of USTCSCC and SCCAS.